# New Superconducting $RbFe_2As_2$: A First-principles Investigation

M. Aftabuzzaman and A.K.M.A. Islam[1]

Department of Physics, Rajshahi University, Rajshahi-6205, Bangladesh

**Abstract**

$RbFe_2As_2$ has recently been reported to be a bulk superconductor with $T_c$ = 2.6 K in the undoped state, in contrast to undoped $BaFe_2As_2$ with a magnetic ground state. We present here the results of the first-principles calculations of the structural, elastic and electronic properties for this newest superconductor and discuss its behaviour in relation to other related systems.

**1. Introduction**

The discovery in 2008 of superconductivity at $T_c$ = 26 K in $LaO_{1-x}F_xFeAs$ [1] has triggered intensive research on the superconductivity of iron pnictides. On replacing a La atom with other rare-earth atoms or by introducing oxygen vacancies [2, 3] $T_c$ value has already exceeded 55 K, which has thus opened a new avenue for high-$T_c$ material research besides cuprates.

The ternary iron arsenide $AFe_2As_2$ (A = alkaline earth element) is another well known structure type having tetragonal $ThCr_2Si_2$-type structure (space group $I4/mmm$) [4] which provides very similar conditions. These iron-arsenide compounds $AFe_2As_2$ are known to become superconducting with $T_c$'s up to 38 K upon alkali metal substitution for the A element [5-7], or partial transition metal substitution for Fe [8]. Sasmal *et al.* [6] have synthesized some of these new high-$T_c$ Fe-based superconducting compounds, $AFe_2As_2$ (A = K, Cs, K/Sr and Cs/Sr). The compounds $AFe_2As_2$ enter a spin-density-wave state (SDW) with increasing electron number (Sr-content) [6]. The compounds represent p-type analogues of the n-doped rare-earth oxypnictide superconductors. In contrast to undoped $BaFe_2As_2$ with a magnetic ground state, superconductivity with relatively low $T_c$'s was reported in the undoped $KFe_2As_2$ ($T_c$ = 3.8 K) and $CsFe_2As_2$ ($T_c$ = 2.6 K) [6]. The $T_c$-value increases with partial substitution of Sr for K and Cs and peaks at 37 K for $x$ = 0.5-0.6 Sr substitution. $AFe_2As_2$ (A= K and Cs) crystallize in the $ThCr_2Si_2$ structure type [9, 10]. The system has identical $(Fe_2As_2)$ layers as in ROFeAs, but separated by single elemental A (Rb in this case) layers. In stacking the $(Fe_2As_2)$ layers in $AFe_2As_2$, the layers are oriented such that the As-As distances between adjacent layers are closest. Nevertheless, interlayer As-As distances in $AFe_2As_2$ are effectively nonbonding. In ROFeAs, adjacent $(Fe_2As_2)$ layers are stacked parallel, with identical orientations, and the $(Fe_2As_2)$ layers are further isolated by more complex $(La_2O_2)$ slabs.

The stimulating endeavours have continued for further works on synthesis, studies of structural and electromagnetic properties, and simulation of similar other materials [11]. The alkali metal iron arsenide $RbFe_2As_2$ with the heavier element Rb also exists [10], but its physical properties have not been reported so far [12,13]. Very recently Bukowski *et al.* [13] reported the occurrence of superconductivity in the undoped $RbFe_2As_2$. These authors synthesized polycrystalline samples of RbFe$_2$As$_2$ in two steps. First, RbAs and Fe$_2$As$_2$ were prepared from pure elements in evacuated and sealed silica tubes. Then, appropriate amounts of RbAs and Fe$_2$As$_2$ were mixed, pressed into pellets and annealed at 650 $^0$C for several days in evacuated and sealed silica ampoules. Powder X-ray diffraction analysis revealed, that the synthesized RbFe$_2$As$_2$ is single phase material which is a bulk superconductor with $T_c$ = 2.6 K [13]. According to the authors as a new representative of iron pnictide superconductors, superconducting RbFe$_2$As$_2$ contrasts with BaFe$_2$As$_2$, where the Fermi level is higher and a magnetic instability is observed. Thus, the solid solution series (Rb,Ba)Fe$_2$As$_2$ is a promising system to study the cross-over from superconductivity to magnetism.

---

[1] Corresponding author: azi46@ru.ac.bd (A.K.M.A. Islam)

The electronic and structural behaviour have already demonstrated the crucial role of the (Fe$_2$As$_2$)-layers in the superconductivity of the Fe-based layered systems. The additional special feature of having simple elemental A layers may provide new avenues to superconductivity, provided we understand their behaviour. In this paper, first-principles calculations of the structural, elastic and electronic properties for the newest RbFe$_2$As$_2$ superconductor are made and the results discussed in comparison with other related systems.

## 2. Computational Methods

The *ab-initio* calculations were performed using the CASTEP program [14]. The geometrical optimization was done for RbFe$_2$As$_2$ (space group 139 *I*4/*mmm*) treating system as metallic with density mixing treatment of electrons. The generalized gradient approximation (GGA) of Perdew, Burke, and Ernzerhof (PBE) [15] potentials have been incorporated for the simulation. We have used a 12×12×3 Monkhorst grid to sample the Brillouin zone. All structures have been fully optimized until internal stress and forces on each atom are negligible. For all relevant calculations the plane wave basis set cut-off used is 330 eV and the convergence criterion is 0.5×10$^{-5}$ eV/atom.

## 3. Results and Discussions

### 3.1 Geometrical optimization

The optimized lattice parameters, internal coordinates, some interatomic distance and bond angles As-Fe-As are shown in the Table 1 along with those obtained by Powder X-ray diffraction analysis by Bukowski *et al*. [13]. We also include in the table the structural data for Ba$_x$K$_{1-x}$Fe$_2$As$_2$ (x = 0, 0.6) [5, 16] for comparison purpose. The optimized lattice parameters obtained by us are in close agreement with the observed values due to Bukowski *et al*. [13]. The crystal structure of RbFe$_2$As$_2$ (ThCr$_2$Si$_2$-type, space group *I*4/*mmm*) is shown in Fig. 1 along with the structure of ROFeAs (space group *P*4/*nmm*) for comparison.

Table 1. Structural parameters of RbFe$_2$As$_2$ and Ba$_x$K$_{1-x}$Fe$_2$As$_2$(x = 0,0.4).

| Compound | $a$ (Å) | $c$ (Å) | $V$ (Å$^3$) | $z$ | $d_{(Rb/Ba-As)}$ (Å) | $d_{(Fe-As)}$ (Å) | $d_{(Fe-Fe)}$ (Å) | Bond angles As-Fe-As (deg.) |
|---|---|---|---|---|---|---|---|---|
| RbFe$_2$As$_2$[This] | 3.7818 | 14.5254 | 207.74 | 0.3401 | 3.542 | 2.300 | 2.674 | 110.6, 108.9 |
| RbFe$_2$As$_2$[13]* | 3.863 | 14.447 | 215.59 | | | | | |
| BaFe$_2$As$_2$[16] | 3.9625 | 13.0168 | 204.38 | 0.3545 | 3.382 | 2.403 | 2.802 | 111.1, 108.7 |
| Ba$_{0.6}$K$_{0.4}$Fe$_2$As$_2$[5] | 3.9170 | 13.2968 | 204.01 | 0.3538 | 3.384 | 2.396 | 2.770 | 109.2, 109.4 |

*Exptl.

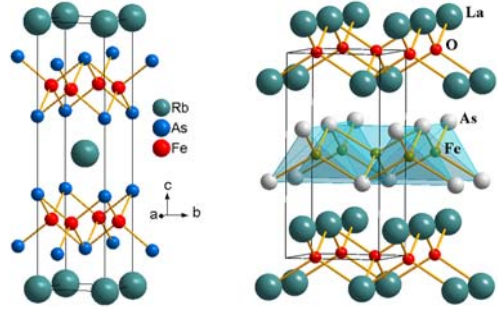

Fig. 1. (Colour online) Crystal structure of RbFe$_2$As$_2$ (ThCr$_2$Si$_2$-type, space group *I*4/*mmm*). For comparison the structure of ROFeAs (space group *P*4/*nmm*) is also shown (*Right*).

## 3.2 Mechanical Characteristics - Elastic Properties

The mechanical characteristics, for example, the elastic properties of the new material RbFe$_2$As$_2$ is of interest because they relate to such fundamental solid state phenomena as specific heat, Debye temperature, and Grüneisen parameter. There are a few empirical dependences $T_c$ and mechanical characteristics of superconductors which are known to exist [17, 18], according to which high values of $T_c$ belong, as a rule, to materials with high compressibility (small bulk modulus $B$). In fact, bulk moduli turn out to be no higher than 200 GPa for many systems with increased $T_c$. We have calculated the six independent elastic constants obtained by the finite strain techniques for the body centred tetragonal RbFe$_2$As$_2$. Table 2 shows these values along with the results for SrFe$_2$As$_2$ reported by Ivanovskii [11]. In order to check the mechanical stability, the values of the elastic stiffness constants will now be discussed. For a crystal with tetragonal symmetry, the mechanical stability is governed by the following Born's conditions (see [11]):

$$C_{11} > 0, \quad C_{33} > 0, \quad C_{44} > 0, \quad C_{66} > 0, \quad (C_{11} - C_{12}) > 0,$$

$$(C_{11} + C_{33} - 2C_{13}) > 0 \quad \text{and} \quad [2(C_{11} + C_{12}) + C_{33} + 4C_{13}] > 0.$$

Obviously the calculated elastic constants of RbFe$_2$As$_2$ are all positive and satisfy the mechanical stability criteria for a tetragonal crystal. On the other hand one of the above conditions is not satisfied for SrFe$_2$As$_2$.

Table 2. Elastic constants, $C_{ij}$ for monocrystalline RbFe$_2$As$_2$.

| Compound | $C_{11}$ | $C_{33}$ | $C_{44}$ | $C_{66}$ | $C_{12}$ | $C_{13}$ | $C_{16}$ |
|---|---|---|---|---|---|---|---|
| RbFe$_2$As$_2$ [This] | 96.9 | 80.4 | 28.6 | 30.1 | 37.7 | 46.2 | 0 |
| SrFe$_2$As$_2$ [11] | 166.1 | 65.0 | ~0 | 80.5 | 30.2 | 36.9 | - |

The theoretical polycrystalline elastic moduli for RbFe$_2$As$_2$ may be calculated from the set of independent elastic constants. Hill [19] proved that the Voigt and Reuss equations represent upper and lower limits of the true polycrystalline constants. He showed that the polycrystalline moduli are the arithmetic mean values of the moduli in the Voigt ($B_R$, $G_R$) and Reuss ($B_R$, $V_R$) approximation, and are thus given by $B_H \equiv B = \frac{1}{2}(B_R + B_V)$ and $G_H \equiv G = \frac{1}{2}(G_R + G_V)$. The expression for Reuss and Voigt moduli can be found elsewhere (see [20]). The Young's modulus Y and Poisson's ratio $\nu$ are then computed from these values using the following relationship: $Y = 9BG/(3B+G)$, $\nu = (3B-Y)/6B$. The calculated elastic parameters (bulk moduli *B*, compressibility *β*, shear moduli *G*, Young's moduli *Y* and the Poisson ratio *ν*) for polycrystalline $RbFe_2As_2$ are shown in Table 3. We compare the polycrystalline elastic moduli of $RbFe_2As_2$ with those of $SrFe_2As_2$ [11]. It is found that $RbFe_2As_2$, like $SrFe_2As_2$, is found to be a soft material. This is evident from the value of relatively smaller bulk moduli for $RbFe_2As_2$ (~59 GPa), which is smaller than the bulk moduli (122-210 GPa) of other well known superconducting materials such as $MgB_2$, $MgCNi_3$, YBCO and $YNi_2B_2C$ [21-24].

Table 3. Calculated elastic parameters for polycrystalline $RbFe_2As_2$.

|  | *B* (GPa) | *β* (GPa$^{-1}$) | *G* (GPa) | *Y* (GPa) | *ν* |
|---|---|---|---|---|---|
| $RbFe_2As_2$[This] | 59.2 | 0.01688 | 26.6 | 69.4 | 0.31 |
| $SrFe_2As_2$ [11] | 61.7 | 0.01621 | 32.1* (2.3) | 82.1* (6.8) | 0.28* (0.42) |

*The reported calculated values (shown within brackets) are in error. We estimated *G* value (in Voigt approximation) and hence *Y*, *ν* using data in ref. [11].

Finally, we can discuss the criterion of brittleness of a material by checking the ratio *B/G*. We know that a material is brittle if *B/G* < 1.75 [25]. According to this $RbFe_2As_2$ lies just above the border of brittleness as *B/G* is 2.23. It is known that the values of the Poisson ratio (*ν*), minimal for covalent materials (*ν* = 0.1), increase for ionic systems [26]. In our case, the value of *ν* for $RbFe_2As_2$ is 0.31, which is indicative of sizable ionic contribution in intra-atomic bonding.

### 3.3 Electronic Band Structure

Figs. 2 (a,b) show the band structures and total and atomic-resolved *l*-projected DOSs in $RbFe_2As_2$. The two lowest bands lying around -12.5 eV below the Fermi level arise mainly from Rb 4*p* and As 4*s* states and are separated from the near-Fermi valence bands by a gap. These valence bands are located in the energy range from -4.9 eV up to the Fermi level $E_F$ = 0 eV and are formed predominantly by As 4*p* and Fe 3*d* states. The corresponding DOS includes three main subbands. The first, second and third subbands range from valence band bottom to -3.5 eV, -3.5 to -1.8 eV and -1.8 to $E_F$, respectively (Fig. 2 b). The first subband is formed As 4*p* and Fe 3*d* states.

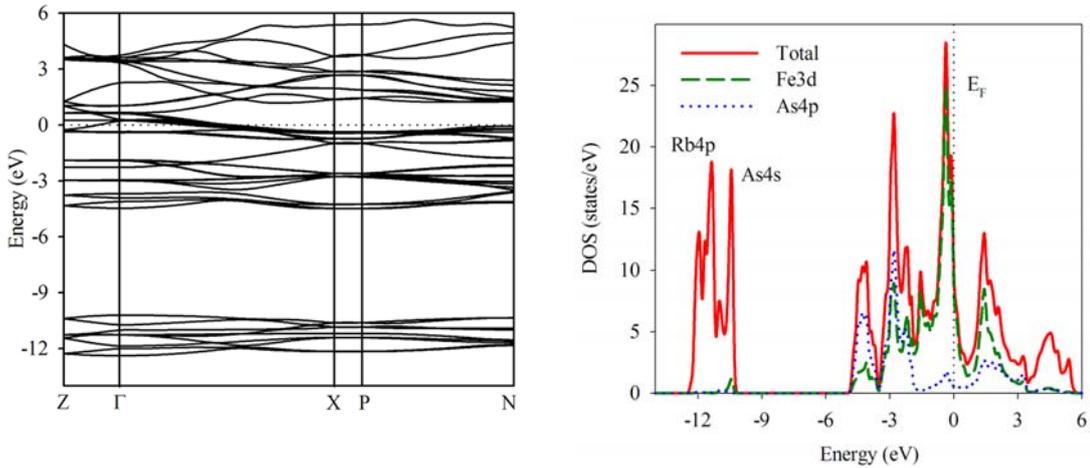

Fig. 2. (a) Electronic band structure and (b) total and partial density of states of $RbFe_2As_2$.

The second subband is derived from As 4$p$ together with an admixture from the Fe 3$d$ states. Thus first two subbands contain strongly hybridized Fe 3$d$ - As 4$p$ states and are responsible for the covalent Fe-As bonding. The third Fe 3$d$-like band intersects the Fermi level and continues to the next unoccupied subband. Hence predominant Fe states with very small admixture of As states prevail near-Fermi region. Thus the DOS region near the Fermi level is formed mainly by the states [FeAs] layers. The contributions from the valence states of Rb to the occupied three subbands and the bottom of the conduction subband are negligible. Thus Rb atoms are in the form of cations Rb$^{1+}$. This shows that, unlike LaOFeAs, the Rb atomic sheets and the [Fe-As] layers are linked by ionic interactions. RbFe$_2$As$_2$ exhibits significant electron deficiency, whereas BaFe$_2$As$_2$ or SrFe$_2$As$_2$ is isoelectronic to LaOFeAs. This may be easily seen by valence counts of $[(Rb)^{1+}]_{0.5}(FeAs)^{0.5-}$ in RbFe$_2$As$_2$ compared to $(LaO)^{1+}(FeAs)^{1-}$ [6]. The electronic and structural behaviour demonstrate the crucial role of the (Fe$_2$As$_2$)-layers in the superconductivity of the Fe-based layered systems, and the special feature of having elemental A layers is expected to provide new path to superconductivity. This system may be described as a quasi-2D ionic metal. The conduction, occurring only on the [Fe-As] layers, is anisotropic.

Now let us discuss some of the electronic behaviours in Ba$_x$K$_{1-x}$Fe$_2$As$_2$ ($x = 0, 0.5$) to show similarities and dissimilarities with those of undoped RbFe$_2$As$_2$. It has been shown by Shein and Ivanovskii [27] that the valence band extends from -5.4 eV to $E_F$ and from -4.9 eV to $E_F$ for BaFe$_2$As$_2$ and Ba$_{0.5}$K$_{0.5}$Fe$_2$As$_2$, respectively. These are derived basically from the Fe 3$d$ states hybridized at the bottom of the valence band with the As 4$p$ states. But there are some dissimilarities in the DOS profiles of BaFe$_2$As$_2$ and Ba$_{0.5}$K$_{0.5}$Fe$_2$As$_2$ which result from the deformations of the [FeAs4] tetrahedra forming the (Fe-As) layers due to the partial replacement of Ba by K [27]. Further, for Ba$_{0.5}$K$_{0.5}$Fe$_2$As$_2$, the admixtures of the Ba and K states in the valence band are absent which indicates that the system remains a quasi-2D ionic metal. The most important difference in the DOS for nonsuperconducting BaFe$_2$As$_2$ and superconducting Ba$_{0.5}$K$_{0.5}$Fe$_2$As$_2$ is the location of the Fermi level [27]. For the hole-doped Ba$_{0.5}$K$_{0.5}$Fe$_2$As$_2$, a decrease in the band filling leads to the movement of the Fermi level in the region of the higher binding energies. As a result, $E_F$ in Ba$_{0.5}$K$_{0.5}$Fe$_2$As$_2$ is shifted downwards and is located on the slope of the sharp peak $C$ (Fig. 2 in ref. [27]) in the region of the enhanced DOS. Thus, the total DOS at the Fermi level for Ba$_{0.5}$K$_{0.5}$Fe$_2$As$_2$ increases by $\sim 20\%$ from the value for BaFe$_2$As$_2$. This enhanced $N(E_F)$ is derived entirely from the Fe 3$d$ states, with negligible contribution from the As 4$p$ states. From what we have discussed it may be said that the appearance of superconductivity in undoped RbFe$_2$As$_2$ is due to the enhanced DOS and the overall shape of its valence DOS (Fig. 2) which is very similar to that of the K-doped BaFe$_2$As$_2$ system.

## 4. Conclusion

We have studied the electronic structure of the newly discovered oxygen-free RbFe$_2$As$_2$ in comparison with similar systems. The density functional theory predicts that RbFe$_2$As$_2$ can be described as a quasi-2D ionic metal. The additional special feature of having simple elemental A layers may provide new ways to superconductivity, provided we understand the behaviour of this type of compounds. Further studies are necessary to understand the mechanism of superconducting coupling for this type of systems.